\documentclass{article}
\usepackage{amsmath,amsthm,amssymb}
\usepackage{a4wide}

\begin{document}

\title{Comment on `Revisiting the Exact Dynamical Structure Factor of the Heisenberg Antiferromagnetic Model' \\ by A. H. Bougourzi}

\author{Jean-S\'ebastien Caux\\
%
Institute for Theoretical Physics, University of Amsterdam, \\
Science Park 904, Postbus 94485, \\
1090 GL Amsterdam, The Netherlands
}

\maketitle

\begin{abstract}
We point out the erroneous reasoning and disprove the conclusions contained in a recent preprint by A. H. Bougourzi ({\sf arxiv:}1402.3855v1) concerning the spin structure factor of the Heisenberg model at zero field in the thermodynamic limit, as calculated using the vertex operator approach.
\end{abstract}

\vspace{10mm}
In a recent manuscript \cite{2014_Bougourzis_piece_of_shit}, much existing work using the vertex operator approach for the calculation of correlation functions of the Heisenberg model in zero field and in the thermodynamic limit is criticized. In particular, it is claimed that the earlier results concerning two-spinon \cite{1997_Karbach_PRB_55} and four-spinon \cite{2006_Caux_JSTAT_P12013} contributions to the dynamical structure factor are incorrect. The criticism focuses on an alleged failure to treat delta functions correctly when expressing the structure factor as a summation over intermediate states. 

\subsubsection*{Handling Dirac delta functions}
Quoting the text after equation (24) in \cite{2014_Bougourzis_piece_of_shit}, {\it `Now a source of an error that plugged [sic] all existing papers dealing with the exact DSF in that they all got the normalization overall factor incorrect is due to the tricky step of expressing each of following sum ... in terms of the usual Dirac delta function $\delta (x)$.'} The sum in question is the one on the left-hand side of equation (26) of \cite{2014_Bougourzis_piece_of_shit}, which we here reproduce textually:
\begin{eqnarray}
\sum_{m \in \mathbb{Z}} \exp 2 i m (k + \sum_{l=1}^n p (\beta_l) ) \stackrel{?}{=} 2\pi \sum_{l' = -n/2 + 1}^0 \delta (2 (k + \sum_{l=1}^n p (\beta_l) - 2\pi l')) \nonumber \\
= \pi \sum_{l' = -n/2 + 1}^0 \delta ( k + \sum_{l=1}^n p (\beta_l) - 2\pi l'), \hspace{10mm}
 -n\pi \leq p(\beta_1) + p (\beta_2) \leq 0.
\label{eq26}
\end{eqnarray}
In the above, $p(\beta)$ are individual spinon momenta varying in the interval $]-\pi, 0]$. The last line should of course be corrected to read $-n\pi < \sum_{l=1}^n p (\beta_l) \leq 0$. 

More importantly, an attentive reader will have noticed that a parenthesis in the right-hand side of the first line is misplaced. The correct version (with changed parenthesis pointed to by the arrow) of the right-hand side is the Dirac comb
\begin{eqnarray}
\sum_{m \in \mathbb{Z}} \exp 2 i m (k + \sum_{l=1}^n p (\beta_l) )
= 2\pi \sum_{l' \in \mathbb{Z}} \delta (2 {\boldsymbol(}k + \sum_{l=1}^n p (\beta_l) \!\stackrel{\downarrow}{\boldsymbol)} - 2\pi l').
\end{eqnarray}
We can rewrite this (using the standard scaling property of the delta function) as two separate sums
\begin{equation}
\pi \sum_{l' \in \mathbb{Z}} \delta (k + \sum_{l=1}^n p (\beta_l) - \pi l' )
= \pi \sum_{l' \in \mathbb{Z}} \sum_{\alpha=0,\pi} \delta ( k + \sum_{l=1}^n p (\beta_l) + \alpha - 2\pi l').
\label{correctdeltas}
\end{equation}
For $n$ spinons (only $n$ even is relevant since there are always even numbers of spinons in our discussion), only $n/2$ values of $l'$ are relevant since $\sum_{l=1}^n p(\beta_l)$ takes values in a range of width $n\pi$. 
This $n/2$ is `factorized' in the right-hand side of equation (27) of \cite{2014_Bougourzis_piece_of_shit} by shifting spinon momenta by units of $2\pi$. The $\alpha = \pi$ terms are however completely missing, which means that half the delta functions are missing.
This is the first relevant mistake. However, in equation (23), which is special to the isotropic limit, there is a factor $1 + e^{i(k+ \sum_l p(\beta_l))}$ which is simply put to zero by the extra $\alpha=\pi$ term which we found above and which was overlooked in \cite{2014_Bougourzis_piece_of_shit}. It would thus seem that this error luckily `cancels out' for the isotropic case (though it certainly would not in the anisotropic case), provided equation (23) of \cite{2014_Bougourzis_piece_of_shit} is correct. Is it?

\subsubsection*{Relations between form factors}
Equation (23) in \cite{2014_Bougourzis_piece_of_shit} is based on equation (18) in the same manuscript, which identifies in the first lines the following two form factors (which we write for convenience in terms of the $\xi$ variables generically used in \cite{JimboBOOK}, although this relation should be understood to only be valid in the isotropic limit):
\begin{equation}
{}_i\langle 0 | \sigma^+_0 | \xi_n, ..., \xi_1 \rangle_{\epsilon_n, ..., \epsilon_1; i} \stackrel{?}{=} -~ {}_{1-i}\langle 0 | \sigma^+_0 | \xi_n, ..., \xi_1 \rangle_{\epsilon_n, ..., \epsilon_1; 1-i}.
\label{FFrelwrong}
\end{equation}
Using equations (A.12) and (7.16) of \cite{JimboBOOK} and taking for example $\xi_n \rightarrow -\xi_n$, we would conclude from equation (\ref{FFrelwrong}) that 
\begin{equation}
{}_i\langle 0 | \sigma^+_0 | -\xi_n, \xi_{n-1}, ..., \xi_1 \rangle_{\epsilon_n, ..., \epsilon_1; i} \stackrel{?}{=} +~{}_{1-i}\langle 0 | \sigma^+_0 | -\xi_n, \xi_{n-1}, ..., \xi_1 \rangle_{\epsilon_n, ..., \epsilon_1; 1-i}
\end{equation}
which contradicts equation (\ref{FFrelwrong}). The properties of the form factors are instead such that the sign in their ratio depends on where the $\xi$ are sitting, and we have
\begin{equation}
\frac{{}_i\langle 0 | \sigma^+_0 | \xi_n, \xi_{n-1}, ..., \xi_1 \rangle_{\epsilon_n, ..., \epsilon_1; i}}{{}_{1-i}\langle 0 | \sigma^+_0 | \xi_n, \xi_{n-1}, ..., \xi_1 \rangle_{\epsilon_n, ..., \epsilon_1; 1-i}} = -\frac{{}_i\langle 0 | \sigma^+_0 | -\xi_n, \xi_{n-1}, ..., \xi_1 \rangle_{\epsilon_n, ..., \epsilon_1; i}}{{}_{1-i}\langle 0 | \sigma^+_0 | -\xi_n, \xi_{n-1}, ..., \xi_1 \rangle_{\epsilon_n, ..., \epsilon_1; 1-i}}
\label{FFrelright}
\end{equation}
The sign used in equation (\ref{FFrelwrong}) is thus dependent on the domains in which the $\xi_1, ..., \xi_n$ sit, which was not taken into account in \cite{2014_Bougourzis_piece_of_shit}. 

\subsubsection*{Using the resolution of the identity operator correctly}
One important subtlety of the vertex operator approach which can easily lead to mistakes is that one must be careful about using the resolution of the identity correctly. This point was extensively discussed in references \cite{2008_Caux_JSTAT_P08006} and \cite{2012_Caux_JSTAT_P01007} which were overlooked in \cite{2014_Bougourzis_piece_of_shit}. Namely: the resolution of the identity in \cite{JimboBOOK} (equation (7.21)) reads
\begin{equation}
{\bf 1} = \sum_{j=0,1} \sum_{n \geq 0} \sum_{\varepsilon_n, ..., \varepsilon_1} \frac{1}{n!} \oint \frac{d\xi_n}{2\pi i \xi_n} ... \oint \frac{d\xi_1}{2\pi i \xi_1} ~| \xi_n, ..., \xi_1 \rangle_{ \varepsilon_n, ..., \varepsilon_1; j} ~{}_{j; \varepsilon_1, ..., \varepsilon_n} \langle \xi_1, ..., \xi_n|
\label{JimboMiwa1}
\end{equation}
where the integrations over $\xi$ are taken over the unit circle. The subtle point is that this integral takes the spinon momenta outside of their fundamental domain, and here $p(\xi) \in~ ]-3\pi/2, \pi/2]$. But {\it one must identify states corresponding to $\xi$ and $-\xi$}: equations (A.12), (7.16) and (7.17) of \cite{JimboBOOK} explicitly show that we must perform this identification, and the orthogonality relation on p.104 of \cite{JimboBOOK} (with $\delta (\xi^2/{\xi'}^2)$) confirms it. Equation (\ref{JimboMiwa1}) must thus be used carefully if one integrates over the whole unit circle of $\xi$. As explained in \cite{2008_Caux_JSTAT_P08006} (see equation (19) in there, and its specialization to the gapless case in equation (3.1) of \cite{2012_Caux_JSTAT_P01007}; by the way, that version was for the even-spinon number sector only, and we here give its fully general form), a risk-free version of the resolution of the identity is to integrate the variable $\xi^2$ over the unit circle,
\begin{equation}
{\bf 1} = \sum_{j=0,1} \sum_{n \geq 0} \sum_{\varepsilon_n, ..., \varepsilon_1} \frac{1}{n!} \oint \frac{d\xi^2_n}{2\pi i \xi^2_n} ... \oint \frac{d\xi^2_1}{2\pi i \xi^2_1} ~| \xi_n, ..., \xi_1 \rangle_{ \varepsilon_n, ..., \varepsilon_1; j} ~{}_{j; \varepsilon_1, ..., \varepsilon_n} \langle \xi_1, ..., \xi_n|.
\label{JSC1}
\end{equation}
This guarantees that the spinon momenta are integrated over their proper interval $]-\pi, 0]$ unlike in equation (\ref{JimboMiwa1}), where the spinon momenta are integrated over $]-3\pi/2, \pi/2]$. Using representation (\ref{JSC1}) instead of (\ref{JimboMiwa1}) means that the main mistakes of \cite{2014_Bougourzis_piece_of_shit} are then automatically avoided. This is what has been used in \cite{2006_Caux_JSTAT_P12013} for the isotropic chain and in \cite{2008_Caux_JSTAT_P08006,2012_Caux_JSTAT_P01007} for the anisotropic cases (respectively transverse structure factor of the gapped chain and longitudinal structure factor of the gapless chain), which by the way connect completely smoothly respectively with the Ising and $XY$ limits, as well as with the previously established isotropic chain results \cite{1997_Karbach_PRB_55}, but not with the results of \cite{2014_Bougourzis_piece_of_shit}.

\subsubsection*{Putting things together (in)correctly}
The mistake with delta functions thus comes back to haunt the reasonings of \cite{2014_Bougourzis_piece_of_shit}. Equation (23) is wrong, because the domain of integration corresponds to that used in representation (\ref{JimboMiwa1}) of the resolution of identity rather than that in (\ref{JSC1}), and the term $1 + e^{i(k+\sum_l p(\beta_l))}$ transforms to $1 - e^{i(k+\sum_l p(\beta_l))}$ in half of the integration region because of (\ref{FFrelright}). The omission of the $\pi$-shifted delta functions, together with the incorrect analytic continuation of form factors to their proper values when the spinon momenta go outside their fundamental domain $p \in~ ]-\pi, 0]$, finally compounded by the use of the representation (\ref{JimboMiwa1}) instead of the pitfall-free (\ref{JSC1}), means that wrong results are thus generated for any number of spinons. 

Leaving aside the misunderstanding of Brillouin zone folding permeating \cite{2014_Bougourzis_piece_of_shit}, the main point we wish to make here is that all the mistakes listed above simply lead to an erroneous factor of $1/2$ because the omitted contributions ($\alpha = \pi$ cases in equation (\ref{correctdeltas})) take precisely the same form as the $\alpha = 0$ ones. The discrepancy in the two-spinon result of \cite{2014_Bougourzis_piece_of_shit} is thus fully explained, and the correct two-spinon contribution of $72.89\%$ from \cite{1997_Karbach_PRB_55} once more confirmed. For the four-spinon case, correcting for the above-explained factor of $1/2$ would mean that the predicted four-spinon contribution of \cite{2014_Bougourzis_piece_of_shit}, if its numerics could be trusted, would become $36\%-40\%$, which is inconsistent with the two-spinon figure.

\subsubsection*{Conclusions} 
The conclusions of \cite{2014_Bougourzis_piece_of_shit} have been shown to be erroneous, being based on a conjuction of three mistakes, one elementary, two rather more subtle. All previously-published results on this issue, by the current author or others, stand.

\paragraph{Acknowledgements} I would like to thank Michael Brockmann for substantially contributing to all aspects of this comment (and especially for finding the misplaced parenthesis), as well as Isaac P{\'e}rez Castillo, Davide Fioretto, Michael Karbach and Robert Weston for useful discussions.

\bibliographystyle{unsrt}
\bibliography{/Users/jscaux/WORK/BIBTEX_LIBRARY/BIBTEX_LIBRARY_JSCaux_PAPERS,/Users/jscaux/WORK/BIBTEX_LIBRARY/BIBTEX_LIBRARY_JSCaux_BOOKS,/Users/jscaux/WORK/BIBTEX_LIBRARY/BIBTEX_LIBRARY_JSCaux_OTHERS,/Users/jscaux/WORK/BIBTEX_LIBRARY/BIBTEX_LIBRARY_JSCaux_OWNPAPERS}

\end{document}